\documentclass[12pt]{article}
\usepackage{authblk}
\usepackage{amsmath,amssymb}
\usepackage{listings}
\usepackage{graphicx}
\usepackage{caption}
\usepackage{subfigure}
\usepackage{MnSymbol}

\newtheorem{theorem}{Theorem}

\makeatletter
\renewcommand\@biblabel[1]{\relax}
\makeatother

\begin{document}

\title{A fast algorithm for maximal propensity score matching\thanks{
This work was supported by the  Russian Foundation for Basic Research, under grants
18-01-00074 and 19-07-00397; and  by the Program for fundamental scientific
research of the SB RAS, No. I.1.3., under grant  0314-2019-0008.
}}
\author[]{Pavel S. Ruzankin\footnote{email: ruzankin@math.nsc.ru}}
\affil[]{\normalsize Sobolev Institute of Mathematics, Novosibirsk, Russia\\
Novosibirsk State University, Novosibirsk, Russia}
\date{}
\maketitle
\begin{abstract}
We present a new algorithm which detects the maximal possible number of matched disjoint pairs satisfying a given caliper when a bipartite matching is done with respect to a scalar index (e.g., propensity score), and constructs a corresponding matching.
Variable width calipers are compatible with the technique, provided that the width of the caliper is a Lipschitz function of the index.
If the observations are ordered with respect to the index then the matching needs $O(N)$ operations, where $N$ is the total number of subjects to be matched. The case of 1-to-$n$ matching is also considered.

We offer also a new fast algorithm for optimal complete one-to-one matching on a scalar index when the treatment and control groups are of the same size. This allows us to improve greedy nearest neighbor matching on a scalar index.

\end{abstract}

\section*{Introduction}

Propensity score matching (PSM) is a statistical method widely used in medicine, biology, and
sociology. The method is used to reduce bias in inference due to confounding variables,
when random allocation of subjects to the comparison groups  is not possible.
The method can be used instead of multivariable regressions approach or in conjunction with it.
The method was introduced by Rosenbaum and Rubin (1983).
The method is based on the Neyman--Rubin model of casual inference (see Rubin, 1974).
In the model, we have the observable treatment indicators $T_j,\ j=1,...,N,$
and the observable outcomes $Z_j,\ j=1,...,N$, where $N$ is the number of subjects under study.
$T_j=1$ if the $j$-th subject belongs to the treatment group and $T_j=0$
if the $j$-th subject is in the control group. Following the common terminology, we call
the groups which we want to compare, the treatment one and the control one.
The Neyman-Rubin model is often called counterfactual because it
contains the unobservable variables $Z^{(0)}_{j}$ and $Z^{(1)}_{j}$,
which denote the outcomes for the $j$-th subject had the subject been allocated
to the control group or to the treatment group, respectively.
We have $$Z_j = T_j Z^{(1)}_{j} + (1-T_j)Z^{(0)}_{j}.$$
The effect of treatment is defined as
$${\bf E}Z^{(1)}-{\bf E}Z^{(0)}.$$
We also observe the vectors $W_1,...,W_N$ of background variables.
In this model the random vectors $(Z^{(0)}_j, Z^{(1)}_j, T_j, W_j)$ are independent and identically distributed,
but the components in each vector are mutually dependent.
Rosenbaum and Rubin (1983) defined the propensity score as
$$p(w):=P(T=1|W=w)$$
and proved that
\begin{equation}\label{vvtt}
W \upmodels T\ |\ p(W),
\end{equation}
and
$$(Z^{(0)},Z^{(1)}) \upmodels T\ |\ p(W)$$
if
\begin{equation}\label{nounob}
(Z^{(0)},Z^{(1)})\upmodels T\ |\ W,
\end{equation}
where the symbol $\upmodels$ denotes independence, $(Z^{(0)},Z^{(1)},T,W)$
is a vector with the same distribution as of all $(Z^{(0)}_j, Z^{(1)}_j, T_j, W_j)$.
(Condition (\ref{nounob}) is often called the condition of no unmeasured confounders.)
That means that, for a fixed $p(W)$,
the random vectors $W$ and $(Z^{(0)},Z^{(1)})$ are equally distributed for $T=0$ and for $T=1$.
That implies that
\begin{align}\nonumber
{\bf E}Z^{(1)}&-{\bf E}Z^{(0)}\\
&=\int \Big({\bf E}(Z^{(1)}|T=1,p(W)=q)-{\bf E}(Z^{(0)}|T=0,p(W)=q)\Big)
{\bf P}(p(W)\in dq)\nonumber\\
&=\int \Big({\bf E}(Z|T=1,p(W)=q)-{\bf E}(Z|T=0,p(W)=q)\Big)
{\bf P}(p(W)\in dq). \label{rpsmbase}
\end{align}
To estimate the last integral, we need  the observable variables only. The propensity score  is usually estimated with logistic regression.
Relation (\ref{rpsmbase}) is what PSM is based on. The PSM consists in matching pairs of
treated and control subjects, such that
a treated and a control subject in each pair have close
values of propensity score. Basing on relation (\ref{vvtt}),
we suppose that the matched subjects have close distributions of the background variables
for $T=0$ and for $T=1$.
Therefore, it is possible to apply, e.g., statistical tests to the matched observations.

\medskip

The majority of studies that employ PSM use greedy nearest neighbor matching (GNNM) algorithms without or with caliper restriction (see Austin, 2011), the caliper width being constant. The caliper width is the maximal allowed within-pair score distance for matching. GNNM means that we try to match each treated subject to the nearest (in terms of the score distance) yet unmatched control subject. There also exist optimal matching algorithms which
minimize average or maximal within-pair score distance, but their main drawback is their time complexity and complexity in the sense of implementation.
There seem to be no widely available packages implementing optimal matching under a caliper
which can be easily used.

In the present paper we mainly consider matching with caliper, i.e., matching with limiting maximal
within-pair score distance.

In Sec.~1 we introduce the main algorithm which matches the maximal possible number of subjects in one-to-one matching under a caliper. Besides, we present a new algorithm for optimal complete one-to-one matching, which allows us to improve GNNM. In Sec.~2 we generalize the main algorithm to one-to-many matching and describe how GNNM can be improved for one-to-many matching. The simulation comparison of the new algorithms
with GNNM is presented in Sec.~3. Sec.~4 contains the proofs of optimality for the new algorithms. Sec.~5 considers matching with discontinuous caliper width. Sec.~6 contains some proofs for the complexity of GNNM.

\section{One-to-one matching}

We consider matching disjoint pairs of subjects from two groups, which we will call, using common terminology, treated and control subjects. In other words, a control subject can be matched to no more than one treated subject and vice versa. We will consider only one-dimensional distance, such as in propensity score matching, when the distance between subjects is the distance between points on the real line corresponding to these subjects, assuming each subject is somehow projected to a unique point on the real line. We will call these points the scores of the subjects. However no assumptions are made on how these points are related to the subjects.

Let $X_i$, $i=1,...,K$, and $Y_j$, $j=1,...,L$, be the scores of treated and control subjects, $K$ and $L$ being the total numbers of treated and control subjects, respectively. $X_j$ and $Y_j$ may take any values on the real line, not necessarily
on the interval $(0,1)$. Thus, the algorithms below can be used, e.g., for matching
by the logits of propensity scores.
Let $N=K+L.$
Let $c=c(x,y)\ge 0$ be the caliper for our matching, i.e., we match only pairs $(i,j)$ such that $|X_i-Y_j|\le c(X_i,Y_j)$. We assume that the caliper is Lipschitz in both arguments with constants 1, i.e., for all $x,y,t$,
\begin{eqnarray}\label{calex}
|c(x,y)-c(x+t,y)|&\le&|t|,\\
|c(x,y)-c(x,y+t)|&\le&|t|.
\label{caley}
\end{eqnarray}
We will consider discontinuous calipers in Sec.~5.

For a discussion of situations where caliper constraints are important for balancing the matched groups see Rosenbaum (2017) and Austin (2011).
Variable caliper width can be useful in situations when, in some domains of values of the propensity score, there are significantly more controls per a treated subject than in other domains (e.g., see examples in Pimentel et al., 2015b). In such cases we can vary the caliper width depending on the density of the number of controls per a treated subject.

\subsection{The main algorithm}

A natural problem is to find the maximal number of pairs that can be matched under a given caliper.
Though this problem can be solved employing network flow optimization algorithms (e.g., see Hansen and Klopfer, 2006), the known algorithms have complexity not less than $O(N^2)$ if no assumptions on sparsity are made. This approach to matching problems was used, e.g., by Rosenbaum (2012, 2017) and Pimentel et al (2015a, 2015b).

Our main goal is to introduce a fast algorithm detecting the maximal possible number of matched pairs and constructing a corresponding matching. Our algorithm has complexity $O(N)$ when
both the treated and control subjects are sorted with respect to the score:
\begin{equation}\label{sort}
X_1\le X_2\le\cdots\le X_K
\quad\mbox{and}\quad Y_1\le Y_2\le \cdots \le Y_L.
\end{equation}

Thus, once we have sorted the observations (which takes
$O(N \log N)$ operations),
we can reasonably fast
solve the inverse problem of finding the minimal constant caliper suitable
for matching $q$ percent of data for a given $q$ (e.g., we may want to match at least $q=75\%$ of data and wish to find out which minimal caliper $c$ would be sufficient).
For instance,
if the score lies on the segment $[0,1]$ then
$l$ runs of the algorithm ($O(lN)$ operations) yield the accuracy of $2^{-l}$
for the minimal caliper. Indeed, first we can run the algorithm for the caliper $c=0.5$
and if it matches not less than $q$ percent of the data then the needed caliper lies on the segment
$[0,0.5]$ and, at the next step, we run the algorithm for $c=0.25$, otherwise
the needed caliper lies on the interval $(0.5,1]$ and
we take $c=0.75$ for the next step. Repeating the steps sequentially halving the interval,  at step $l$ we obtain the interval of length $2^{-l}$ containing the minimal constant caliper suitable
for matching $q$ percent of the data.

From now on we assume that relation (\ref{sort}) holds, unless nearest neighbor matching is considered.

Let us now introduce the main algorithm. The variable $M$ will contain the current number of matched pairs. After the algorithm finishes, $M$ contains the maximal possible number of matched pairs. $A_m$ and $B_m$ store the index numbers of treated and control subject, respectively, in the $m$-th matched pair.

We present the algorithm as the following pseudocode:

\medskip
\noindent{\bf Algorithm A.}
\begin{lstlisting}
$M:=0$
$i:=1$
$j:=1$
while ($i\le K$ and $j\le L$)
   if ($|X_i-Y_j|\le c(X_i,Y_j)$)
      $M:=M+1$
      $A_M:=i$
      $B_M:=j$
      $i:=i+1$
      $j:=j+1$
   else
      if ($X_i<Y_j$)
         $i:=i+1$
      else
         $j:=j+1$
      end if
   end if
end while
\end{lstlisting}
As we see, the algorithm just walks through all the observations and successively collects all feasible pairs.

The algorithm requires $O(N)$ operations since in each iteration of the while-loop the variable $i$ or $j$ or both are increased. Certainly, to apply the algorithm, first we must sort the observations with respect to the score, which requires $O(N \log N)$ operations.

\begin{theorem}\label{th1}
Algorithm~A produces the maximal possible number of matched pairs under a caliper
satisfying $(\ref{calex})$--$(\ref{caley})$.
\end{theorem}

This theorem is proved in Subsection~4.2. Some simulations for the algorithm are presented in Subsection 3.1.


\subsection{Optimal complete matching and improving nearest neighbor matching}

We will call a one-to-one matching (without replacement) \emph{complete} if
the sizes of the treatment and control groups coincide $K=L$ and all the subjects are matched.
A caliper restriction may prevent some subjects from being matched, therefore
we will consider only complete matchings without caliper.

Colannino et al. (2007) used observations' sorting for complete one-to-one matching on a scalar index (without applying a caliper). Their algorithm's complexity is $O(N)$ after the observations are ordered with respect to the score. That algorithm minimizes the cost of matching
$$\sum_{(i,j)} |X_i-Y_j|,$$
where the sum is taken over all matched pairs $(i,j)$, which is equivalent to minimizing
the average within-pair score distance.

We offer the following  new algorithm of the same complexity, which minimizes that cost along with some other costs, including
maximal within-pair score distance.

\medskip
\noindent {\bf Algorithm~B.} Let $K=L$ and the observations be sorted as in $(\ref{sort})$.
\emph{Match $X_j$ to $Y_j$ for all $j=1,...,K$.}

\begin{theorem}\label{thnn}
Let $K=L$ and the observations be sorted as in $(\ref{sort})$.
Then, among all complete matchings, Algorithm~B minimizes average within-pair score distance
as well as the following cost functions:
maximal within-pair score distance
\begin{equation}\label{maxcost}
\max_{(i,j)} |X_i-Y_j|,
\end{equation}
\begin{equation}\label{phicost}
\sum_{(i,j)} \varphi(X_i-Y_j)
\mbox{ for any convex nonnegative function } \varphi,
\end{equation}
and
\begin{equation}\label{altcost}
\sum_{(i,j)} |X_i-Y_j| h(\max\{|X_i-a|,|Y_j-a|\}),
\end{equation}
where $h$ is a nondecreasing nonnegative continuous function, $a$ is a real number.
The maximum and the sums are taken over all matched pairs $(i,j)$.
\end{theorem}

The theorem is proved in Subsection~4.1.

\medskip
A curious result is that the ``opposite'' matching yields the ``counteroptimal'' cost.
\begin{theorem}\label{thnn2}
Let $K=L$ and the observations be sorted as in $(\ref{sort})$.
Match $X_j$ to $Y_{K-j+1}$ for all $j=1,...,N$. Then, among all complete matchings, this matching
maximizes the costs $(\ref{maxcost})$ and $(\ref{phicost})$, in particular, the average within-pair score distance
is maximized.
\end{theorem}

The theorem is proved in Subsection~4.1.

\medskip
That shows that if one considers matching on a scalar index then the problem of optimal (but not complete) matching minimizing or maximizing (\ref{phicost}) is essentially the problem of choosing the optimal subsets of the observations. After the subsets of treated and control subjects are chosen, it is sufficient just to order the observations.

\medskip
{\bf Improving nearest neighbor matching.} Let us apply Theorem~\ref{thnn}
 to a non-complete one-to-one matching on a scalar index, e.g., GNNM,
under the caliper $c(x,y)$ satisfying (\ref{calex}) and (\ref{caley}).
Let $\tilde X_1,...,\tilde X_{\tilde M}$ and $\tilde Y_1,...,\tilde Y_{\tilde M}$
be the ordered scores of the matched treated and control observations:
\begin{equation}\label{suborder}
\tilde X_1\le \cdots \le\tilde X_{\tilde M},\quad
\tilde Y_1\le \cdots \le \tilde Y_{\tilde M}.
\end{equation}
Then, by Theorem~\ref{th1}, rematching these (matched) observations with Algorithm~A will produce,
under the caliper $c(x,y)$, the maximal possible number of pairs, which is $\tilde M$. Since Algorithm~A goes sequentially through the ordered observations, it will match the observations corresponding to
$\tilde X_j$ and $\tilde Y_j$ for each $j$.
This proves that matching
the observations corresponding to
$\tilde X_j$ and $\tilde Y_j$ for each $j$ obeys the caliper $c(x,y)$.

Such rematching is optimal with respect to average and maximal within-pair distances
by Theorem~\ref{thnn}, and can improve the average and maximal within-pair distances
as is shown by simulations in Subsection~3.1.

In other words, to improve some matching, we can rearrange the pairs of matched observations via ordering the matched observations as in (\ref{suborder}) and then
matching the observations corresponding to
$\tilde X_j$ and $\tilde Y_j$ for each $j$. Such rematching does not break the caliper restriction because of the optimality of Algorithm~A.

\medskip
Note also that GNNM with caliper has complexity similar to that of Algorithm~A (see Sec.~6). If the observations are ordered as in (\ref{sort}) then sequential GNNM has complexity $O(N)$, while for unordered observations GNNM has complexity $O(N \log N)$.

\newpage
\section{$1$-to-$n$ matching}

\subsection{The main algorithm}

Algorithm~A can be modified for $1$-to-$n$ matching. We assume that a treated subject is to be matched to no more than $n$ control subjects, and a control subject must not be matched to more than one treated subject. Some authors call these settings
matching with a varying number of controls (e.g., see Pimentel et al 2015b).

The following pseudocode uses the same variables as above. $D_i$ is  the number of controls matched to the $i$-th treated subject. The variable $k$ corresponds to the current number of controls matched to the $i$-th treated subject.
\medskip


\noindent{\bf Algorithm C.}
\begin{lstlisting}
$M:=0$
$i:=1$
$j:=1$
$k:=0$
$D_i:=0$ for all $i=1,...,K$
while ($i\le K$ and $j\le L$)
   if ($|X_i-Y_j|\le c(X_i,Y_j)$)
      $k:=k+1$
      $M:=M+1$
      $A_M:=i$
      $B_M:=j$
      $D_i:=k$
      if ($k=n$)
         $k:=0$
         $i:=i+1$
      end if
      $j:=j+1$
   else
      if ($X_i<Y_j$)
         $k:=0$
         $i:=i+1$
      else
         $j:=j+1$
      end if
   end if
end while
\end{lstlisting}

The complexity is still $O(N)$ and does not depend on $n$ since, as above, in each iteration of the while-loop the variable $i$ or $j$ or both are increased.

\begin{theorem}\label{thalgb}
Algorithm~C maximizes the number of matched control subjects or, in other words, the number of matched pairs for $1$-to-$n$ matching under a caliper satisfying
$(\ref{calex})$--$(\ref{caley})$.
\end{theorem}

The theorem is proved in Subsection~4.2. Note that the algorithm does not maximize the number of matched treated subjects. Some simulations for the algorithm are presented in Subsection~3.2.

\subsection{Improving nearest neighbor matching}

We can rematch matched observations to improve GNNM for 1-to-$n$ matching.
We will consider the following GNNM scheme. The matching is done in $n$ passes. In each pass,
we try to match each treated subject to only one nearest yet unmatched control.
Such scheme is aimed at increasing the number of matched treated subjects.

We can improve such matching by applying the argument of Subsection~1.2. For this, we consider each pass as a one-to-one matching and, in each pass, we rematch, as in (\ref{suborder}), the observations matched
by GNNM in this pass.

Such rematching does not alter the number of matched controls and the number of matched treated subjects.
The simulation comparison of this algorithm with GNNM and Algorithm~C is in Subsection~3.2.

\section{Simulation comparison with nearest neighbor matching}

\subsection{Simulation for one-to-one matching}

In this subsection we compare Algorithm~A with one-to-one greedy nearest neighbor matching (GNNM) and GNNM followed by the optimal rematching described in Subsection~1.2.
GNNM means that we first try to match match the first treated subject to the nearest control subject, then the second treated subject and so on. The matching is done
without replacement.  All the three algorithms have similar complexities (see Sec.~6).

 We take $X_i$  to be i.i.d. random  variables on the interval $(0,1)$ with the density
 $2y,$ $0<y<1,$ and $Y_j$ to be i.i.d. random  variables on the interval $(0,1)$ with the density
 $2-2y,$ $0<y<1$.
 We use the calipers $c=c_1$ for Algorithm~A and $c=c_2:=0.02$  for GNNM. Each of the following graphs is constructed by 10,000 simulation runs. In each simulation, the treatment group and control group are of the same size of 100 or 1000.

First we try to compare the numbers of matched pairs for the algorithms in the case when $c_1=c_2=0.02$. Fig.~1 depicts the empirical cumulative distribution functions for the numbers of matched pairs. The graphs are plotted for Algorithm~A (solid lines) and GNNM (dashed lines). We see that under these settings Algorithm~A matches more pairs than GNNM. It makes little sense to compare algorithms that match significantly different numbers of pairs. If one algorithm is allowed to match a smaller number of pairs compared to another algorithm then the former algorithm can easily produce lesser maximal and average distances between the scores of paired observations. On the other hand, lesser numbers of pairs lead to less significant p-values and powers for statistical tests applied to matched observations.

\begin{figure}[!h]
    \centering
    \subfigure[$K=L=100,\ c_1=c_2=0.02$]{
        \includegraphics[width=0.45\textwidth]{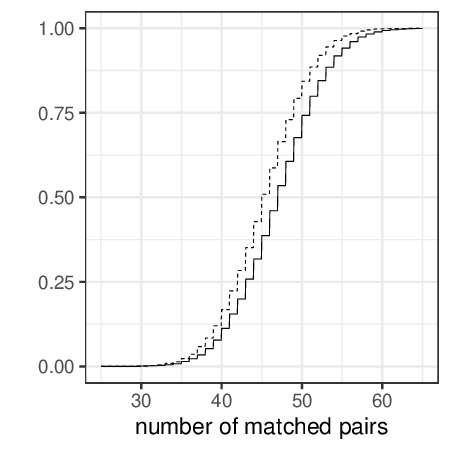}
    }
    \subfigure[$K=L=1000,\ c_1=c_2=0.02$]{
        \includegraphics[width=0.45\textwidth]{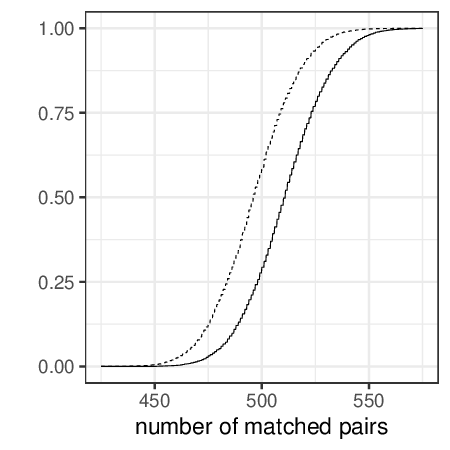}
    }
    \vspace{-12pt}
    \caption{Empirical CDFs for the numbers of pairs matched by the algorithms}\label{fig0}
\end{figure}

\newpage
That is why, for the next graphs, we choose some $c_1<c_2$ to make the numbers of pairs matched by Algorithm~A and GNNM be similar. Fig.~2--4 depict the empirical cumulative distribution functions for the number of matched pairs, the maximal distance between the scores of paired observations, and the average distance between the scores of paired observations, respectively. The graphs are plotted for Algorithm~A (solid lines), GNNM (dashed lines) and GNNM with rematching (\ref{suborder}) (dotted lines).

\begin{figure}[!h]
    \centering
    \subfigure[$K=L=100,\ c_1=0.0155$]{
        \includegraphics[width=0.45\textwidth]{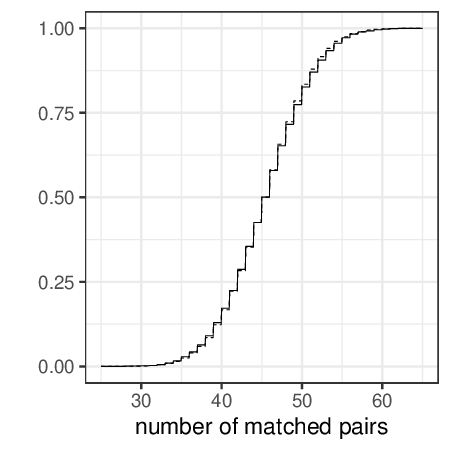}
    }
    \subfigure[$K=L=1000,\ c_1=0.0065$]{
        \includegraphics[width=0.45\textwidth]{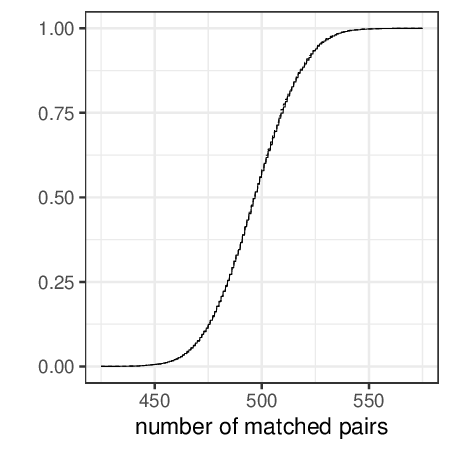}
    }
    \vspace{-12pt}
    \caption{Empirical CDFs for the numbers of pairs matched by the algorithms}\label{fig1}
\end{figure}

\begin{figure}[!h]
    \centering
    \subfigure[$K=L=100,\ c_1=0.0155$]{
        \includegraphics[width=0.45\textwidth]{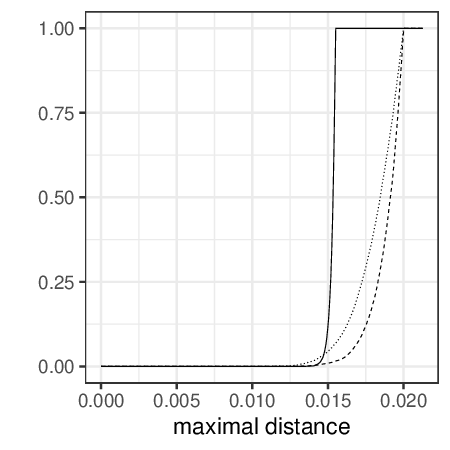}
    }
    \subfigure[$K=L=1000,\ c_1=0.0065$]{
        \includegraphics[width=0.45\textwidth]{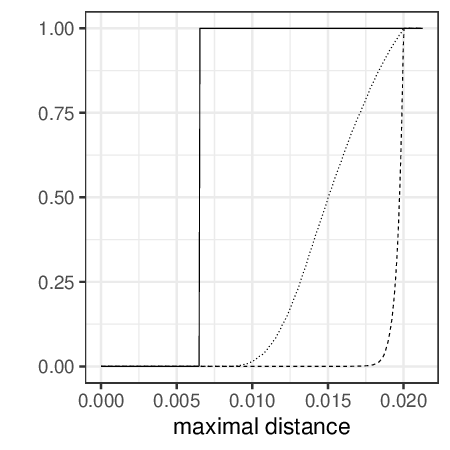}
    }
    \vspace{-12pt}
    \caption{Empirical CDFs for the maximal within-pair score distance}\label{fig2}
\end{figure}

\begin{figure}[!h]
    \centering
    \subfigure[$K=L=100,\ c_1=0.0155$]{
        \includegraphics[width=0.45\textwidth]{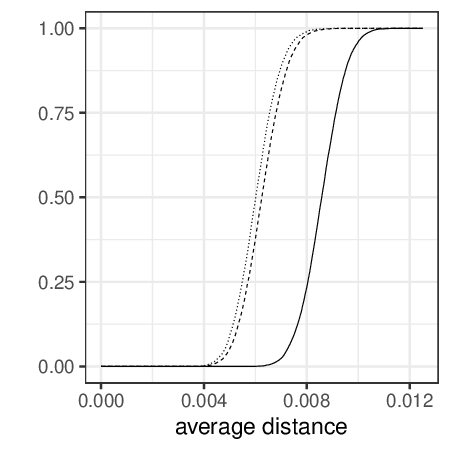}
    }
    \subfigure[$K=L=1000,\ c_1=0.0065$]{
        \includegraphics[width=0.45\textwidth]{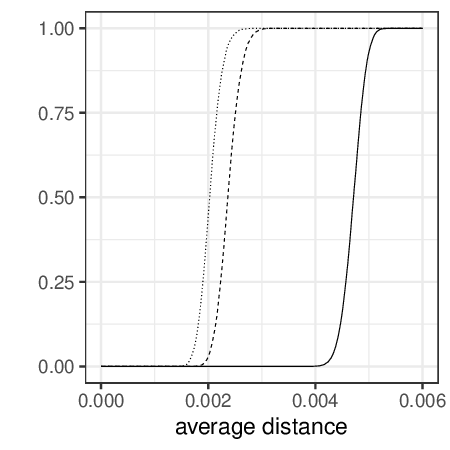}
    }
    \vspace{-12pt}
    \caption{Empirical CDFs for the average within-pair score distance}\label{fig3}
\end{figure}

\newpage
The simulation results for the case $K=L=100$ are summarized in the following table, where the means
for the values plotted on Fig.~2--4 are presented:\\
\begin{tabular}{ l |p{1.5cm}|p{2.5cm}|p{2.5cm} }
  \hfill Mean of:& number of matched pairs& maximal within-pair score~distance & average within-pair score~distance \\
  \hline
  Algorithm~A & 45.5 & 0.0153 & 0.0086\\
  \hline
  GNNM with rematching (\ref{suborder}) & 45.5 & 0.0181 & 0.0060\\
  \hline
  GNNM & 45.5 & 0.0188 & 0.0063\\
\end{tabular}

\medskip
The next table presents the means for the case $K=L=1000$:\\
\begin{tabular}{ l |p{1.5cm}|p{2.5cm}|p{2.5cm} }
  \hfill Mean of:& number of matched pairs& maximal within-pair score~distance & average within-pair score~distance \\
  \hline
  Algorithm~A & 496.8 & 0.0065 & 0.0047\\
  \hline
  GNNM with rematching (\ref{suborder}) & 496.6 & 0.0151 & 0.0020\\
  \hline
  GNNM & 496.6 & 0.0196 & 0.0024 \\
\end{tabular}

\medskip
We see that if we want to minimize the average distance between the scores of paired observations then we may choose GNNM with rematching (\ref{suborder}). But if we want to minimize the maximal distance between the scores in pairs then we may prefer Algorithm~A.

The other argument for choosing Algorithm~A may be its complexity. If we have to match ``big data'', the complexity may be of more importance than the accuracy of matching.

For explicit practical recommendations an extensive simulation comparison may be needed, like that in Austin (2014). 

\subsection{Simulation for 1-to-3 matching}
In this subsection we compare Algorithm~C with GNNM and GNNM
with rematching (\ref{suborder}) for 1-to-3 matching.
GNNM and GNNM with rematching (\ref{suborder}) are accomplished
according to the scheme described in Subsection~2.2.

As in the above simulations, $X_i$  are i.i.d. random  variables on the interval $(0,1)$ with the density
$2y,$ $0<y<1,$ and $Y_j$ are i.i.d. random  variables on the interval $(0,1)$ with the density
$2-2y,$ $0<y<1$. We use the caliper $c=c_1$ for Algorithm~C and $c=c_2:=0.02$  for GNNM.
The estimates below are computed by 10,000 simulation runs. In each simulation, the treatment group is of the size $K=100$ or $K=1000$, and the control group is of the size $L=3K$.

The unweighted average within-pair score distance is computed as the average
distance among all matched pairs. For computing the weighted average within-pair distance,
the weight of each pair is the inverse of the number of controls the current treated subject is matched to. So the sum of the weights of all pairs for a matching is the number of matched treated subjects.

The following table summarizes the simulation results for the case $K=100$, $L=300$,
$c_1=0.0147$:\\
\begin{tabular}{p{2.2cm} |p{1.5cm}|p{1.8cm}|p{1.8cm}|p{2cm}|p{2cm} }
  \hfill Mean of:& number of matched pairs& maximal within-pair score distance & weighted average within-pair score distance&
  unweighted average within-pair score distance&
  number of matched treated subjects \\
  \hline
  Algorithm~C & 141.6 & 0.01462 & 0.00298& 0.00827&58.8\\
  \hline
  GNNM with rematching (\ref{suborder})  & 141.6 & 0.01888 & 0.00267&0.00491&71.6\\
\hline
  GNNM & 141.6 & 0.01941 & 0.00282&0.00507&71.6\\
\end{tabular}

 \medskip
 \newpage
 The following table summarizes the simulation results for the case $K=1000$, $L=3000$,
$c_1=0.0055$:\\
\begin{tabular}{p{2.2cm} |p{1.5cm}|p{1.8cm}|p{1.8cm}|p{2cm}|p{2cm} }
  \hfill Mean of:& number of matched pairs& maximal within-pair score distance & weighted average within-pair score distance&
  unweighted average within-pair score distance&
  number of matched treated subjects \\
  \hline
  Algorithm~C & 1494.2 & 0.00550 & 0.00146& 0.00410&604.5\\
  \hline
  GNNM with rematching (\ref{suborder}) & 1494.1 & 0.01797 &0.00087 &0.00152&748.4\\
\hline
  GNNM & 1494.1 & 0.01983 & 0.00101&0.00170&748.4\\
\end{tabular}

\medskip
For one-to-many matching, one may be interested in maximizing the number of matched treated subjects. If it is the case then GNNM with rematching (\ref{suborder}) is to be preferred. While if one is interested in maximizing the number of matched pairs and minimizing the maximal within-pair score distance then he/she is to choose Algorithm~C.

\section{Proofs of optimality}

\subsection{Proofs for complete matching}

We offer two proofs for Theorem~\ref{thnn}. The first proof is substantially based
on the results on the Monge--Kantorovich mass trasfer problem. The second proof is straightforward.

\medskip\noindent
{\bf The first proof of Theorem~\ref{thnn}.}
Let $P$ and $Q$ be two probability measures on $\mathbb R$ with the cumulative distribution functions $F$ and $G$, respectively. Let $\varphi:{\mathbb R}\to {\mathbb R}$ be a convex
nonnegative function. Then (relation (2.14) in Rachev, 1985)
\begin{equation}\label{mmmtp}
\inf_{\xi\sim P,\ \eta\sim Q}{\bf E}\varphi(\xi-\eta)=
{\bf E}\varphi(\xi_0-\eta_0)=\int_0^1\varphi\big(F^{-1}(t)-G^{-1}(t)\big)dt,
\end{equation}
where the infimum is taken over all random variables $\xi$ and $\eta$
on a common probability space with the distributions
$P$ and $Q$, respectively,
$$\xi_0=F^{-1}(\omega),\quad \eta_0=G^{-1}(\omega),$$
$\omega$ is a random variable uniformly distributed on $(0,1)$,
$$F^{-1}(t)=\sup\{y\,:\,F(y)<t\}$$
is the quantile transformation of $F$.

We will apply relation (\ref{mmmtp}) to prove the optimality of Algorithm~B.
Take $P$ and $Q$ such that $P(\{X_j\})=Q(\{Y_j\})=1/K$ for all
$j=1,...,K$. Then $\xi_0=X_j$ and $\eta_0=Y_j$ for
$\omega\in((j-1)/K,j/K)$. Therefore,
$$\inf_{\xi\sim P,\ \eta\sim Q}{\bf E}\varphi(\xi-\eta)=
\frac{1}{K}\sum_{j=1}^K \varphi(X_j-Y_j).$$
To prove the optimality (\ref{phicost}) it remains to notice that
$$\inf_{\xi\sim P,\ \eta\sim Q}{\bf E}\varphi(\xi-\eta)\le
 \frac{1}{K} \min_\sigma\sum_{j=1}^K \varphi(X_j-Y_{\sigma(j)}),$$
 where the minimum is taken over all permutations $\sigma$ of the set $1,...,K$. Hence,
$$\min_\sigma\sum_{j=1}^K \varphi(X_j-Y_{\sigma(j)})=\sum_{j=1}^K \varphi(X_j-Y_j).$$


The optimality (\ref{phicost}) is proved. The optimality (\ref{altcost})
can be proved analogously by Example after Theorem 2 in Rachev (1985).

Let us now prove the optimality (\ref{maxcost}).
If there are two complete one-to-one matchings ${\cal M}_1$ and ${\cal M}_2$ of $N=2K$ subjects, such that
$$\max_{(i,j)\in {\cal M}_1} |X_i-Y_j| < \max_{(i,j)\in {\cal M}_2} |X_i-Y_j|$$
 then there exits a $\gamma>1$ such that
$$\sum_{(i,j)\in {\cal M}_1} |X_i-Y_j|^\gamma < \sum_{(i,j)\in {\cal M}_2} |X_i-Y_j|^\gamma.$$
Indeed, it is sufficient to take a $\gamma>1$ such that
$$K\left(\max_{(i,j)\in {\cal M}_1} |X_i-Y_j|\right)^\gamma
<\left(\max_{(i,j)\in {\cal M}_2} |X_i-Y_j|\right)^\gamma.$$ Hence, since the function
$\varphi(y)=|y|^\gamma$ is convex for any $\gamma\ge 1$ and thus (\ref{phicost}) is minimized for any such $\varphi(y)$, the functional $\max_{(i,j)} |X_i-Y_j|$ is minimized as well.

The theorem is proved.


\medskip\noindent
{\bf The second proof of Theorem~\ref{thnn}.}
Let us prove the optimality ({\ref{phicost}).
Let a complete matching $\mathcal M$ match $X_i$ to $Y_j$ and $X_k$ to $Y_l$, where
$i<k$ but $j>l$. Thus, $X_i\le X_k$ but $Y_j\ge Y_l$. Hence, we have
$$ X_i-Y_j\le X_i-Y_l\le X_k-Y_l$$
and
$$ X_i-Y_j\le X_k-Y_j\le X_k-Y_l.$$
Therefore, we have
$$\max\{|X_i-Y_l|,|X_k-Y_j|\}\le\max\{|X_i-Y_j|,|X_k-Y_l|\}
$$
and, since the function $\varphi$ is convex and $(X_i-Y_l)-(X_i-Y_j)=(X_k-Y_l)-(X_k-Y_j)$,
$$\varphi(X_i-Y_j)+\varphi(X_k-Y_l)\ge \varphi(X_i-Y_l)+\varphi(X_k-Y_j).$$
Hence, replacing in $\mathcal M$ the pairs $(i,j)$ and $(k,l)$ by
the pairs $(i,l)$ and $(j,k)$ does not increase the costs (\ref{maxcost}) and (\ref{phicost}).

Sequentially replacing pairs $(i,j)$ and $(k,l)$ such that $i<k$ and $j>l$
by the pairs $(i,l)$ and $(j,k)$, we will finally come to the matching of Algorithm~B. (We may use, e.g., bubble sort scheme for it.) That proves that
the costs (\ref{maxcost}) and (\ref{phicost}) for Algorithm~B are not greater than those  of any other compelte matching.

The optimalities (\ref{maxcost}) and (\ref{phicost}) are proved.

Here we omit the proof of optimality (\ref{altcost}) for the sake of brevity.
 One may see the first proof of the theorem for the proof of that optimality.

The theorem is proved.

\medskip
\noindent {\bf Proof of Theorem~\ref{thnn2}}
can be done analogously to the first proof of
Theorem~\ref{thnn} by virtue of the relation ((2.14) in Rachev, 1985)
$$\sup_{\xi\sim P,\ \eta\sim Q}{\bf E}\varphi(\xi-\eta)=
\int_0^1\varphi\big(F^{-1}(t)-G^{-1}(1-t)\big)dt,$$
where $P$, $Q$, $F$, $G$, $\xi$, $\eta$ are the same as in the proof of Theorem~\ref{thnn}.

There is also a straightforward proof of Theorem~\ref{thnn2} analogous to the second
proof of Theorem~\ref{thnn}.

\subsection{Proofs for Algorithms A and C}

For a constant caliper, Theorems~\ref{th1} and \ref{thalgb} can be proved analogously to the first proof of Theorem~\ref{thnn} using the corresponding results on the Monge--Kantorovich
problem in Ruzankin (2001). Here we offer the proofs that are valid for variable width calipers as well.

\medskip
\noindent {\bf Proof of Theorem~\ref{th1}.}
There exists a matching
$\mathcal M$ satisfying the caliper (i.e.,
$|X_i-Y_j|\le c(X_i,Y_j)$ for all $(i,j)\in\mathcal M$) and containing the maximal number of matched pairs.

Consider the first step. If $X_1<Y_1-c(X_1,Y_1)$ then
$$X_1< Y_1-c(X_1,Y_1)+\big(Y_j-Y_1+c(X_1,Y_1)-c(X_1,Y_j)\big)
\equiv Y_j-c(X_1,Y_j)$$
for all $j$, since $Y_j-Y_1+c(X_1,Y_1)-c(X_1,Y_j)\ge 0 $ by (\ref{caley}), and, hence, the first treated subject cannot be used for matching. Analogously
if $Y_1<X_1-c(X_1,Y_1)$ then the first control subject is not suitable for matching by
(\ref{calex}).
Thus first steps of the algorithm skip the observations that cannot be used for matching.

After the above operation we can assume, for the sake of convenience, that $|X_1-Y_1|\le c(X_1,Y_1)$. Let us show that matching now the first treated with the first control subject, as the algorithm does, does not reduce the maximal number of matched pairs, if we match the maximal number of pairs for the remaining $2,\dots,K$-th treated
 and $2,\dots,L$-th control  subjects.

If the first treated or the first control subject are not matched in $\mathcal M$ then removing from $\mathcal M$ a possible pair with the first treated or the first control subject and then
adding $(1,1)$ to $\mathcal M$ does not change the number of pairs in $\mathcal M$. Thus, in this case, matching the pair $(1,1)$
and then matching the maximal number of pairs for the $2,\dots,K$-th treated and $2,\dots,L$-th control subjects yields
the total maximal number of matched pairs.

The case when $\mathcal M$ contains the pair $(1,1)$ is clear.

It remains to consider the case when $\mathcal M$ contains some pairs $(1,j_1)$ and $(i_1,1)$, where $i_1\ne 1$ and $j_1\ne 1$.
In this case we have $X_{i_1} \le Y_1+c(X_{i_1}, Y_1)$ and
$ Y_{j_1}\le X_1+c( X_1,Y_{j_1})$. Therefore
\begin{eqnarray*}
X_{i_1}-Y_{j_1}&\le& Y_1+c(X_{i_1}, Y_1)-Y_{j_1}\\
&=& c(X_{i_1},Y_{j_1})-\big(Y_{j_1}-Y_1+c(X_{i_1},Y_{j_1})-c(X_{i_1}, Y_1)\big) \\
&\le& c(X_{i_1},Y_{j_1})
\end{eqnarray*}
by (\ref{caley})
and analogously
$Y_{j_1}-X_{i_1}\le c(X_{i_1},Y_{j_1})$ by (\ref{calex}). Hence,
$$|X_{i_1}-Y_{j_1}|\le c(X_{i_1},Y_{j_1}).$$
Thus, removing from $\mathcal M$ the pairs $(1,j_1)$ and $(i_1,1)$
and adding the pairs $(1,1)$ and $(i_1,j_1)$ obeys the caliper restriction and does not change the number of pairs in $\mathcal M$.
Again, matching the pair $(1,1)$
and then matching the maximal number of pairs for the $2,\dots,K$-th treated and $2,\dots,L$-th control subjects yields
the total maximal number of matched pairs.

Applying the above argument to the remaining
$2,\dots,K$-th treated
and $2,\dots,L$-th control observations
proves the optimality of Algorithm~A by induction.

The theorem is proved.

\medskip
\noindent{\bf Proof of Theorem~\ref{thalgb}.}
For the case of $1$-to-$n$ matching it suffices
to consider Algorithm~C as Algorithm~A applied to observations where we take $n$ identical treated subjects instead
of each corresponding treated subject from the original observations, i.e., we ``repeat'' each treated subject $n$ times.

\section{The case of piecewise Lipschitz caliper}

In this section we will describe an algorithm which yields a maximal number of pairs
under somewhat weaker conditions on the caliper. We will consider one-to-one matching though it is easy to modify the algorithm below for the case of $1$-to-$n$ matching just like it was done for Algorithm~A.

We will assume that, first, the caliper
is ``Lipschitz-nondecreasing'':
\begin{eqnarray}\label{calexnd}
c(x,y+t)&\ge &c(x,y)-t\ \mbox{ for all }t>0,x,y,\\
c(x+t,y)&\ge &c(x,y)-t\ \mbox{ for all }t>0,x,y
\label{caleynd}
\end{eqnarray}
and, second, the caliper is piecewise Lipschitz in both arguments: there exist
disjoint intervals $[a_1,a_2),...,[a_{U-1},a_U)$ covering the domain of $X_i$
and disjoint intervals $[b_1,b_2),...,[b_{V-1},b_V)$ covering the domain of $Y_j$
such that, for each $u=1,...,U-1$,
\begin{eqnarray}\label{calexp}
c(x+t,y)&\le &c(x,y)+t\ \mbox{ for all }
x,x+t\in [a_u,a_{u+1}),\ t>0,\ y
\end{eqnarray}
and, for each $v=1,...,V-1$,
\begin{eqnarray}\label{caleyp}
c(x,y+t)&\le &c(x,y)+t\ \mbox{ for all }
y,y+t\in [b_v,b_{v+1}),\ t>0,\ x.
\end{eqnarray}

For example, if $c(x,y)=f(x)+g(y)$ or $c(x,y)=f(x)\,g(y)$, where $f(x)$ and $g(y)$ are nondecreasing nonnegative step functions,
or if $c(x,y)=e^{-|x|}(1-(y-\lfloor y\rfloor))$, where $\lfloor y \rfloor $ denotes the greatest integer not greater than $y$,
then conditions (\ref{calexnd})--(\ref{caleyp}) are satisfied.

Let us now introduce an algorithm for a caliper satisfying (\ref{calexnd})--(\ref{caleyp}).
As above, $M$ is the current number of matched pairs. After the algorithm finishes, $M$ is the maximal number of matched pairs. $A_m$ and $B_m$ store the index numbers of treated and control subject, respectively, in the $m$-th matched pair.

$I_1,...,I_U$ are increasing numbers such that $X_i\in [a_u,a_{u+1})$ whenever
$I_u\le i < I_{u+1}$; and increasing numbers $J_1,...,J_V$ are such that $Y_j\in [b_v,b_{v+1})$ whenever
$J_v\le j < J_{v+1}$.
Computing $I_1,...,I_U$ and  $J_1,...,J_V$ given $a_1,...,a_U$, $b_1,...,b_V$,
$X_1,...,X_K$, and $Y_1,...,Y_L$ requires $O(N)$ operations.
If some of the intervals $[a_u,a_{u+1})$ contain no observations $X_i$
then we are to take the number of intervals $[I_u, I_{u+1})$ lesser than
the number of intervals $[a_u,a_{u+1})$, but, to simplify notations, we use the same
$U$ to enumerate $I_u$, $u=1,...,U$. The same is done for
the intervals $[J_v, J_{v+1})$.

For each $u=1,...,U-1$, we will have $S_u=i$ if and only if $i\in [I_u,I_{u+1}]$, the observations $X_{I_u},...,X_{i-1}$
are already matched or discarded, and either $i$ equals $I_{u+1}$ or $X_i$ is currently neither matched nor discarded.
Symmetrically, for each $v=1,...,V-1$, we will have $T_v=j$ if and only if $j\in [J_v,J_{v+1}]$, the observations $Y_{J_v},...,Y_{j-1}$
are already matched or discarded, and either $j$ equals $I_{u+1}$ or $Y_j$ is currently neither matched nor discarded.

We will assume that ``and'' in the if-statement means that the second condition is checked only if the first one is true.

\medskip
\noindent{\bf Algorithm D.}
\begin{lstlisting}
$M:=0$
$S_u:=I_u$ for all $u=1,...,U$
$T_v:=J_v$ for all $v=1,...,V$
$i:=1$
$j:=1$
$u0:=1$
$v0:=1$

function $increment\_i$()
   $S_{u0}:=S_{u0}+1$
   $i:=i+1$
   if ($i=I_{u0+1}$)
      $u1:=u0+1$
      while ($u1<U$ and $S_{u1}=I_{u1+1}$) $u1:=u1+1$
      $u0:=u1$
      $i:=S_{u0}$
   end if
end function

function $increment\_j$()
   $T_{v0}:=T_{v0}+1$
   $j:=j+1$
   if ($j=J_{v0+1}$)
      $v1:=v0+1$
      while ($v1<V$ and $T_{v1}=J_{v1+1}$) $v1:=v1+1$
      $v0:=v1$
      $j:=T_{v0}$
   end if
end function
\end{lstlisting}
\newpage
\begin{lstlisting}
while ($i\le K$ and $j\le L$)
   if ($X_i<Y_j$)
      for ($v=v0,...,V-1$)
         if ($T_v<J_{v+1}$ and $|X_i-Y_{T_v}|\le c(X_i,Y_{T_v})$)
            $M:=M+1$
            $A_M:=i$
            $B_M:=T_v$
            $increment\_i$()
            if ($v0=v$)
               $increment\_j$()
            else
               $T_v:=T_v+1$
            end if
            next while
         end if
      end for
      $increment\_i$()
   else
      for ($u=u0,...,U-1$)
         if ($S_u<I_{u+1}$ and $|X_{S_u}-Y_j|\le c(X_{S_u},Y_j)$)
            $M:=M+1$
            $A_M:=S_u$
            $B_M:=j$
            $increment\_j$()
            if ($u0=u$)
               $increment\_i$()
            else
               $S_u:=S_u+1$
            end if
            next while
         end if
      end for
      $increment\_j$()
   end if
end while
\end{lstlisting}

The complexity of the last algorithm is $O((U+V)N)$ since each iteration of the while-loop requires $O(U+V)$ operations and in each iteration the variable $i$ or $j$ or both are increased.

The proof for the maximality of the number of matched pairs almost repeats the proof of Theorem~\ref{thalgb} for Algorithm~C. The main difference that if, say, at some step $X_i<Y_j$ then we have to check sequentially whether $X_i$ can be matched to each group $\{Y_{J_v},...,Y_{J_{v+1}-1}\}$, $v=1,...,V-1$, of the control observations. As above, by (\ref{caleyp}) it is sufficient for each group to check whether $X_i$ can be matched to the first unmatched element of the group.
Relations (\ref{calexnd}) and (\ref{caleynd}) ensure that matching $X_i$ to the first unmatched element of the first suitable group does not diminish the number of matched pairs below its maximal value.

\section{Complexity of nearest neighbor matching}

In this section we discuss the complexity of one-to-one greedy nearest neighbor matching (GNNM) under a caliper. We match sequentially the first treated subject, the second one, and so on. The matching is done without replacement. In this section no assumptions on the caliper are made.

{\bf Nearest neighbor matching for sorted observations.} Let us consider observations sorted as in (\ref{sort}). We want to match the observations by GNNM with the caliper $c(x,y)$.

This can be done in $O(N)$ time if we use a list data structure for the control observations. The list can be organized as the vector containing the controls' scores, and two integer vectors for left and right pointers of the list cells. (In fact, in this case the vector for right pointers is not needed, since we use the right pointers only to move to the right through the list until we meet the first control with the score not less than that of the current treated subject.)

{\bf Nearest neighbor matching for unordered observations.} Now
we make no assumptions on the order of the observations. For instance,
the treated observations may be randomly permuted, the permutations being uniformly distributed. Such a permutation can be done in $O(N)$ time.

The GNNM can be done in $O(N \log N)$ time by the following algorithm.

\medskip
First we build a balanced binary tree for the control observations, which requires
$O(N\log N)$ operations (e.g., see Ruzankin, 2019). Each node of the tree contains the number $j$ of the corresponding control observation. The left subtree of each node contains control observations with the scores lesser than or equal to that of the node, and the right subtree contains controls with the  scores greater than or equal to that of the node.

The main problem in using such trees for matching is in dealing with already matched observations. We offer the following solution to the problem.

In the process of matching, a node becomes void after the corresponding control observation is matched.  The algorithm does not allow a void node to have one or no down edges. So if a leaf node's observation is matched then the node is removed from the tree and, after that, if the parent of this node is a void node then it is also deleted, its up edge and (the only) down edge being ``glued'' together. Analogously, if we match an observation from a node that has only one down edge then the node is removed, its up and down edges being ``glued''. But if we match a control from a node that has two down edges then this node just becomes void, but keeps containing the number of the corresponding observation.

For each treated subject, the algorithm goes down the tree.
At each step of this process there are two to four guesses. Each guess has its score and is the number of a node or a dummy guess. Each guess is flagged as static or branching. Each step transforms the guesses or, when the guesses cannot be transformed, tries to match an observation from the guesses.
For the first step we take the root node as a branching guess, and two static dummy guesses with some scores $p_1<\min_j Y_j$ and $p_3>\max_j Y_j$, respectively.

Let, at some step, we have $2\le l\le 4$ guesses with the scores $p_1\le\cdots\le p_l$ for matching a treated observation with the score $X$. First we select from these guesses the left and right guesses.
If $p_j\le X\le p_{j+1}$ for some $j$ then the $j$-th and $(j+1)$-th guesses are assigned to be the left and right guess, respectively. Note that we always have $p_1\le X\le p_l$.

If both the left and right guesses are static then we match the one of them closest to $X$,
satisfying the caliper, and being not dummy, if any, and then proceed to matching the next treated subject.

Next, if the left or right guess is static then it reproduces itself as a static guess for the next step.

If the left or right guess is a void node then it puts its both children to be branching guesses for the next step. Note that a void guess cannot be static.

If the left guess is not void and branching then it reproduces itself as a static guess for the next step and puts its right child (if any) as a branching guess for the next step.

If the right guess is not void and branching then it reproduces itself as a static guess for the next step and puts its left child (if any) as a branching guess for the next step.

Thus we have two to four guesses prepared for the next step and can proceed to it.

\medskip
As we see, for each treated subject, we need $O(\log N)$ operations to travel down the tree and select the nearest control neighbor, and then we need $O(1)$ operations to remove the corresponding void nodes. Thus the total complexity is $O(N\log N)$.

\section*{Acknowledgments}
The author is grateful to Ben B. Hansen and Mark M. Fredrickson for the discussion
which has substantially improved the paper. The author thanks the reviewers for useful comments.



\begin{thebibliography}{99}


\bibitem{austin2}
Austin, P. C. (2011), ``An Introduction to Propensity Score Methods for Reducing the Effects of Confounding in Observational Studies'',
\emph{Multivariate Behavioral Research}, 46, No. 3: Propensity Score Analysis, 399--424.


\bibitem{austin}
Austin, P. C. (2014), ``A comparison of 12 algorithms for matching on the propensity score'', \emph{Statist. Med.}, 33, 1057--1069.

\bibitem{col}
Colannino, J., Damian, M., Hurtado, F. et al.  (2007),
``Efficient Many-To-Many Point Matching in One Dimension'',
\emph{Graphs and Combinatorics}, 23(Suppl 1), 169--178.


\bibitem{hansen}
Hansen, B.~B. and Klopfer, S.~O. (2006), ``Optimal full matching
and related designs via network flows'', \emph{Journal of Computational
and Graphical Statistics}, 15, No.3, 609--627.

\bibitem{p1}
Pimentel, S. D., Kelz, R. R., Silber, J. H, and Rosenbaum, P. R.
(2015a),
``Large, Sparse Optimal Matching With Refined Covariate Balance in an Observational Study of the Health Outcomes Produced by New Surgeons,'' \emph{Journal of the American Statistical Association,} 110, No. 510, 517--527.

\bibitem{p2}
Pimentel, S. D., Yoon, F., and Keele, L. (2015b), ``Variable-ratio matching with fine balance in a study of the Peer Health Exchange,''
\emph{Statistics in Medicine,} 34, No. 30, 4070--4082.



\bibitem{rachev}
Rachev, S. T. (1985),
``The Monge-Kantorovich Mass Transference Problem and Its Stochastic Applications'',
\emph{Theory Probab. Appl.}, 29, No. 4, 647--676.



\bibitem{r1}
Rosenbaum, P. R. (2012), ``Optimal Matching of an Optimally Chosen Subset in Observational Studies,'' \emph{Journal of Computational and Graphical Statistics,} 21, No. 1, 57--71.

\bibitem{r2}
Rosenbaum, P. R. (2017). ``Imposing minimax and quantile constraints on optimal matching in observational studies,'' \emph{Journal of Computational and Graphical Statistics}, 26, No. 1, 66--78.


\bibitem{rr}
Rosenbaum, P. R., Rubin, D. B. (1983), ``The Central Role of the Propensity Score in Observational Studies for Causal Effects,'' \emph{Biometrika,} 70, No. 1, 41--55.


\bibitem{rubin}
Rubin, D. B. (1974).  ``Estimating Causal Effects of Treatments in Randomized and
Nonrandomized Studies''. \emph{Journal of Educational Psychology}, 66, No. 5, 688--701.


\bibitem{ruz}
Ruzankin, P. S. (2001), ``Construction of the optimal joint distribution of two random variables,'' \emph{Theory Probab. Appl.,} 46, No.2, 316--334.

\bibitem{ruz2}
Ruzankin, P. S. (2019), ``A fast algorithm for constructing balanced binary search trees,'' Preprint, https://arxiv.org/abs/1902.02499

\end{thebibliography}
\end{document}